\documentclass[letterpaper,aps,prl,twocolumn,showpacs,preprintnumbers,superscriptaddress]{revtex4}

\usepackage{graphicx}
\usepackage{epsfig}
\usepackage{dcolumn}
\usepackage{bm}
\usepackage{amsmath}
\usepackage{amssymb}
\usepackage{latexsym}
\usepackage[usenames,dvipsnames]{xcolor}
\usepackage{color}
\usepackage{multirow}

\def\({\left(}
\def\){\right)}
\def\<{\left<}
\def\>{\right>}
\def\pl{\partial}

\begin{document}

\title{The circular Kardar-Parisi-Zhang equation as an inflating,
  self-avoiding ring polymer}

\author{Silvia N.\ Santalla}
\affiliation{Departamento de F\'{\i}sica and Grupo Interdisciplinar de
  Sistemas Complejos (GISC), Universidad Carlos III de Madrid, Legan\'es, Spain}
\author{Javier Rodr\'{\i}guez-Laguna}
\affiliation{ICFO--Institut de Ci\`encies Fot\`oniques, Castelldefels,
  Spain}
\affiliation{Departamento de Matem\'{a}ticas and GISC, Universidad
  Carlos III de Madrid, Legan\'{e}s, Spain}
\author{Rodolfo Cuerno}
\affiliation{Departamento de Matem\'{a}ticas and GISC, Universidad
  Carlos III de Madrid, Legan\'{e}s, Spain}

\date{September 5, 2013}

\begin{abstract}
We consider the Kardar-Parisi-Zhang (KPZ) equation for a circular interface
in two dimensions, unconstrained by the standard small-slopes and
no-overhang approximations. Numerical simulations using an adaptive
scheme allow us to elucidate the complete time evolution as a crossover between a 
short-time regime with the interface fluctuations
of a {\em self-avoiding ring} or 2D vesicle, and a long-time regime
governed by the Tracy-Widom distribution expected for this geometry.
For small noise amplitudes, scaling behavior is only of the latter type. Large
noise is also seen to {\em renormalize} the bare physical parameters of the ring,
akin to analogous parameter renormalization for equilibrium 3D membranes.
Our results bear particular importance on the relation between relevant universality
classes of scale-invariant systems in two dimensions.
\end{abstract}


\maketitle

Recently, statistical fluctuations are revealing interesting features
for a number of one-dimensional systems confined to circular
geometries. For instance, for semiflexible polymers of a fixed length,
like constrained DNA rings, the closure condition influences the scaling, shape, and transport
behavior \cite{witz:2011}. Topology is
actually expected to play a key role in a large number of related biophysical
processes, such as e.g.\ translocation in
nanochannels or nanopores \cite{micheletti:2012,luo:2008}, or knot
localization \cite{ercolini:2007}. In particular, circular DNA
molecules in two dimensions have been experimentally found
\cite{witz:2011} to be well described as pressurized vesicles
\cite{leibler:1987}, their scaling behavior depending on the geometry
\cite{camacho:1990}. Thus, for deflated rings (negative pressure
difference, $\Delta p$), fluctuations are in the universality class of
lattice animals, while for $\Delta p=0$, statistics are those of a
ring self-avoiding walk (SAW) \cite{witz:2011}. The latter is
important as a paradigmatic model of polymers
\cite{des_cloizeaux:book} and because the SAW is believed to constitute a
conformally invariant system in two dimensions \cite{cardy_et_al}.

For planar rings evolving far from equilibrium, the closure condition
is also proving non-trivial, as recently observed in experiments with
droplets of turbulent liquid crystals \cite{takeuchi_et_al}, for the
edge of a drying colloidal suspension \cite{yunker:2013,nicoli:2013}, and for many
more systems, from epitaxy to bacterial growth \cite{saito:2012}.
Thus, as proposed in \cite{praehofer:2000}, the probability distribution
function (pdf) of the height fluctuations for interfaces which, like many of these, belong to the
Kardar-Parisi-Zhang (KPZ) universality class \cite{barabasi:book,krug:1997},
depends on the global curvature. The eponymous equation \cite{kardar:1986},
which is the prime representative for these systems, is a continuum
model for the evolution of a rough interface between a (stable,
e.g.\ solid) phase that grows at the expense of a(n unstable,
e.g.\ vapor) phase,
\begin{equation}
\partial_t h = v+ \nu \nabla^2 h + \frac{\lambda}{2} (\nabla h)^2 + \eta(\mathbf{x},t) ,
\label{kpz}
\end{equation}
where $h(\mathbf{x},t)$ is the height field above substrate position
$\mathbf{x} \in \mathbb{R}^d$ at time $t$, $\eta$ is Gaussian white
noise, $v$ is the growth speed for a flat interface, and $\nu>0$,
$\lambda$ are additional parameters. On the one hand, from the point of view of the theory of stochastic
processes, the KPZ equation features a remarkable example of a time crossover \cite{corwin:2012,calabrese}
between the two main universality classes of kinetic roughening \cite{barabasi:book,krug:1997}, namely, the Edwards-Wilkinson (EW) class at short times, and the KPZ class at long times. Experimentally \cite{takeuchi_et_al},
however, while such a crossover may have been seen in $d=1$ for interfaces with a null global curvature,
it has not for the circular geometry case. On the other hand, for such ring-shaped interfaces
height statistics are indeed distinctively described \cite{praehofer:2000}
by the Tracy-Widom (TW) pdf for the largest-eigenvalue of large random matrices in the Gaussian unitary ensemble (GUE),
as recently supported by exact solutions of the KPZ equation on an infinite substrate and a wedge initial condition (but without explicit closure), or for related systems \cite{sasamoto:2010_etal}. This actually occurs
with a remarkable degree of universality \cite{kriecherbauer:2010,corwin:2012},
as the same pdf, critical exponents, and limiting processes apply to discrete models,
continuum equations \cite{alves:2011_etal}, and experiments
\cite{takeuchi_et_al,yunker:2013,nicoli:2013}. Hence, in two dimensions (2D) both pressurized
vesicles and the KPZ equation notably demonstrate the non-trivial role of geometry,
as a part of the universality class and related renormalization-group fixed point
\cite{camacho:1990}, in and out of equilibrium, respectively.


Note, Eq.\ \eqref{kpz} is just the small-slope, single-valued
approximation of a more general equation \cite{rodriguez-laguna:2011},
\begin{equation}
\partial_t \mathbf{r} = \left(A_0 + A_1 K(\mathbf{r}) + A_n \eta(\mathbf{r},t) \right) \mathbf{u}_n ,
\label{ckpz}
\end{equation}
where $\mathbf{r}(t)\in\mathbb{R}^{d+1}$ gives the interface position,
$K(\mathbf{r})$ is the local extrinsic surface curvature,
$\mathbf{u}_n$ is the normal direction pointing towards the unstable
phase, and constants $A_0$, $A_1$, and $A_n$ relate to parameters in
Eq.\ \eqref{kpz} in a simple way \cite{rodriguez-laguna:2011}. They
account for, respectively, the average growth speed along the local
normal direction, surface tension effects, and noise in the local
growth velocity, precisely the physical mechanisms at play in the
formulation of the KPZ equation as a continuum interface model
\cite{kardar:1986}. However, those produced by Eq.\ \eqref{ckpz} are
not constrained to small slopes or lack of overhangs
\cite{rodriguez-laguna:2011}.

For a ring geometry and $d=1$, Eq.\ \eqref{kpz} actually has to be
given up in favor of Eq.\ \eqref{ckpz}, since the closure condition
hinders description of the interface profile by a single-valued
function altogether. Alternative formulations to Eq.\ \eqref{ckpz} are
available, see e.g.\ in \cite{saito:2012}, although most include the
neglect of overhangs, and/or additional simplifications. A natural
question is then whether Eqs.\ \eqref{kpz} and \eqref{ckpz} have the
same dynamic scaling properties. Here we show that, for planar rings,
this is {\em not} the case. Namely, while asymptotics are indeed of the
expected TW-GUE type also for Eq.\ \eqref{ckpz}, which implements
explicitly a closure condition, the early times differ substantially
as compared with Eq.\ \eqref{kpz}: now, for small noise amplitudes
no scaling behavior other than KPZ is obtained, as in experiments \cite{takeuchi_et_al},
while 2D-SAW universality is obtained at short times for large noise amplitude values $A_n$.
Such large fluctuations renormalize additional parameters like $A_1$,
in a form that is reminiscent of surface tension renormalization by
non-equilibrium fluctuations, as experimentally assessed e.g.\ in \cite{el_alaoui_faris:09}
for 3D active membranes. In parallel with the equilibrium behavior
of 2D vesicles \cite{camacho:1990} as a function of $\Delta p$, the change
from early-time SAW to late-time KPZ scaling behavior correlates with an evolution in time from a
freely fluctuating ring to an average circular shape. Thus, the generalized KPZ equation
\eqref{ckpz} also predicts a crossover to occur during the time evolution of the system
between two equally celebrated universality classes under large noise conditions. From this point
of view the experiments in \cite{takeuchi_et_al} correspond to a small-noise condition, while 
a prediction is provided for suitable large noise situations, which should be amenable to 
experimental verification.

\begin{figure}[!t]
\includegraphics[angle=270,width=0.5\textwidth]{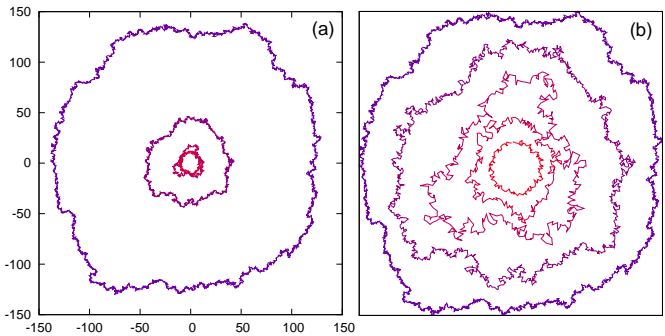}
\caption{(Color online) (a) Interface evolution for $R_0 =10$, $A_0=0.01$,
  $A_1=0.01$, and $A_n=1$, with $\Delta t=0.1$, $l_{\rm min}=0.1$, and
  $l_{\rm max}=1$. Curves for times $t=2,20,150,1000$, and 5000, inner
  to outer. (b) Rescaled view to ease comparison. All units are
  arbitrary.}
\label{fig:1}
\end{figure}

We have performed numerical simulations of Eq.\ \eqref{ckpz} using
planar rings of various initial radii $R_0$ and center $(0,0)$ as
initial conditions. We employ an adaptive algorithm as in
\cite{rodriguez-laguna:2011,blinnikov:1996}, which does not need to
assume a single-valued polar function. The interface is represented by
a chain of points (a ``polymer'') $\{P_i\}_{i=1}^{N(t)}\subset
\mathbb{R}^2$ defining a piecewise continuous curve which always
leaves the stable phase on its left. The distance between them is
forced to remain in an interval $[l_{\rm min}, l_{\rm max}]$,
which is achieved by inserting or removing points
dynamically. Interface properties, like curvature, are evaluated in a
geometrically natural way \cite{rodriguez-laguna:2011}. Unavoidably,
self-intersections occur along the evolution. We always remove the smaller
interface component, eliminating both cavities and
outgrowths, thus implementing self-avoidance and rendering the dynamics
irreversible. This approximation is akin to restricting dynamics to that of
the active zone in growth systems \cite{rodriguez-laguna:2011}. Time updates are via an
Euler-Maruyama scheme with spacing $\Delta t$, sufficiently small that
it does not appreciably influence results.

A set of representative snapshots are shown in Fig.\ \ref{fig:1}, for
$R_0=10$, $A_0=0.01$, $A_1=0.01$, $A_n=1$, and different
times. Qualitatively, the ring can be seen to undergo two different
regimes: (I) for $t\lesssim 100$ it fluctuates without significant
growth while, its shape becoming less and less circular; (II) for
$t\gtrsim 400$, the ring grows steadily, progressively recovering an
average circular shape. In order to interpret these observations, we can
consider the deterministic case, i.e., $A_n=0$. Rings with smaller
$R_0$ than a certain threshold shrink, since the constant average
velocity $A_0$ is not able to compensate for the effect of surface
tension $A_1$. On the other hand, for $R_0\gtrsim A_1/A_0$, the ring
grows very slowly at first, and with velocity $A_0$ for longer
times.

\begin{figure}[!t]
\centering
\includegraphics[angle=270,width=0.5\textwidth]{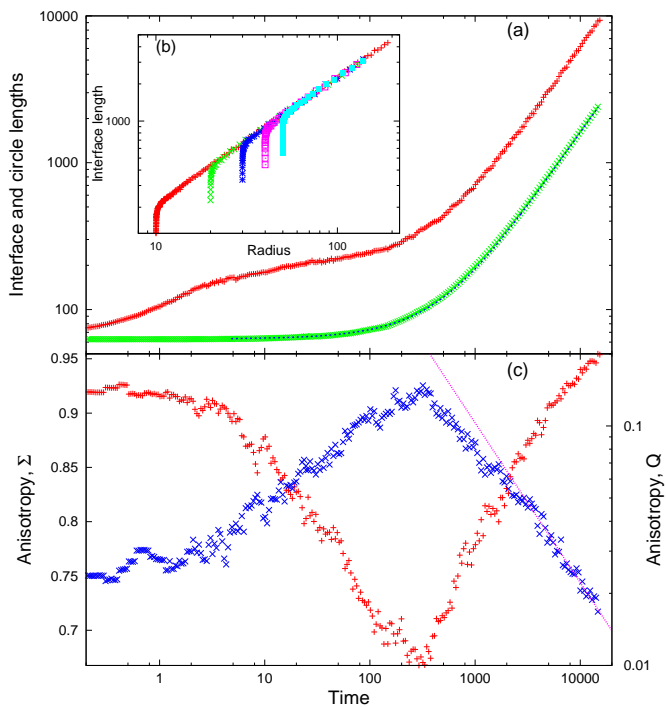}
\caption{(Color online) Evolution of {\em interface shape} for the case shown in
  Fig.\ \ref{fig:1}. (a) Interface length (above) and circle length
  $2\pi R(t)$ (below), with $R(t)$ the fitted radius, vs.\ time. For
  long times, both are linear in $t$. Dashed line: fit to a
  renormalized deterministic growth. (b) Interface length
  vs. radius for different initial radii, $R_0=10,20,30,40$, and $50$
  (growing upwards). (c) Anisotropy vs.\ time: $\Sigma$ ($+$),
  left vertical axis, and quadrupole moment $Q$ ($\times$), right
  vertical axis. The slope of the straight line, $-0.667$, corresponds
  to a fit for long times. Units are arbitrary.}
\label{fig:2}
\end{figure}

From Eq.\ \eqref{ckpz}, a simplified evolution equation can be derived for the average
ring radius $R(t)$,
\begin{equation}
\frac{{\rm d}R(t)}{{\rm d} t} = \tilde{A}_0 + \frac{\tilde{A}_1}{R(t)} ,
\label{R_mf}
\end{equation}
where local variations in the normal velocity are neglected and the total
ring length $L(t)$ is approximated by that of the average circle, see details in \cite{sm}.
Here, $\tilde{A}_{0,1}$ have values that will in general differ from their ``bare''
counterparts, $A_{0,1}$, due to noise-induced renormalization. Fig.\ \ref{fig:2} shows $R(t)$ for the
same parameter choice of Fig.\ \ref{fig:1}. Regimes I and II are clearly distinguished in the growth rate.
Remarkably, the numerical $R(t)$ fits the {\em exact} solution of Eq.\ \eqref{R_mf} \cite{sm} for
$\tilde A_0\approx 0.026>0.01=A_0$ and $\tilde A_1\approx 0.1>0.01=A_1$. For small noise amplitudes,
no such noise renormalization occurs, see \cite{sm}. Thinking of $A_0$ as a
{\em pressure difference} that attempts to ``inflate'' the ring
\cite{leibler:1987,camacho:1990}, the effect of $A_n$ can be thought
of as a {\em fluctuation-induced} pressure boost $\tilde A_0$. Alongside, $\tilde A_1$ becomes
an enhanced surface tension, due to the noisy dynamics. Similar
fits are obtained for a wide range of bare parameters.

Fig.\ \ref{fig:2} (a) also depicts the numerical evolution of the actual $L(t)$
(for a space cut-off $l_{\rm min}$). Very early in regime I, while the
average radius remains almost constant, this length increases due to
fluctuations. In regime II, when $R(t)$ grows steadily, $L(t)$
actually becomes proportional to it. This behavior is
appreciated in Fig.\ \ref{fig:2} (b), where the $L(t)$
is plotted vs.\ $R(t)$. There is a threshold total length,
proportional to $R_0$, below which no radial growth occurs, and above
which both measures become proportional. As seen in Fig.\ \ref{fig:1},
prior to regime II noise basically induces loss of the initial
circular symmetry. In Fig.\ \ref{fig:2} (c) we quantify this effect by
plotting the asymmetry parameter \cite{family:1985} $\Sigma(t)=\langle
S^2_{G1}/S^2_{G2}\rangle$, i.e.\ the ratio of the smallest to largest
eigenvalues, $S^2_{G1}, S^2_{G2}$, of the gyration tensor $S$. This is
frequently used to assess polymer classes in terms of self-avoidance,
dimensionality, rigidity, etc.\ \cite{bishop:1986,camacho:1990,alim:2007,witz:2011}. In our
case, $\Sigma(t)$ decreases with time, reaches a minimum value
$\Sigma(t\simeq 400)\simeq 0.68$, and increases back, approaching the
characteristic value of a circular swollen polymer in regime
II. Other measures of anisotropy lead to the same conclusion, see
e.g.\ Fig.\ \ref{fig:2} (c) for the quadrupole moment, $Q^2={\langle
  |x^2-y^2|\rangle / \langle r^2\rangle}$, where $x$, $y$, and $r$ are
relative to the CM of the $\{P_i\}_{i=1}^{N(t)}$ point distribution.

\begin{figure}[!t]
\centering
\includegraphics[angle=270,width=0.5\textwidth]{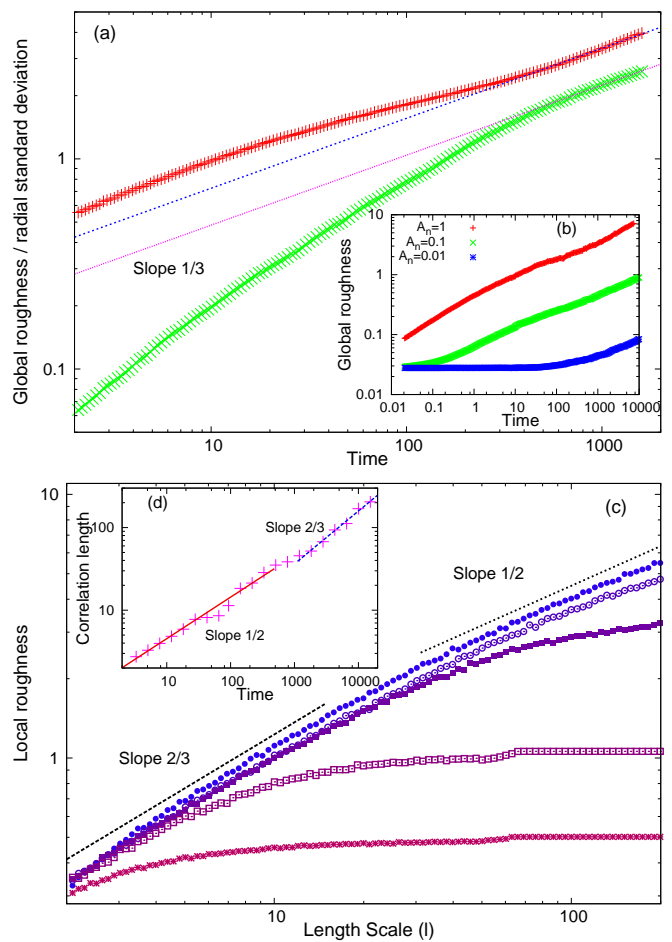}
\caption{(Color online) Evolution of the {\em interface fluctuations} for the case
  shown in Fig.\ \ref{fig:1}. (a) Global roughness ($+$) and standard
  deviation of $R(t)$ ($\times$) vs.\ time. Both straight lines have
  slopes 1/3. (b) $W(t)$ for decreasing noise amplitude, as in legend,
  top (same data as in main panel) to bottom. (c) Local roughness
  vs.\ window size, for $t=2,20,1200,8000$, and $14000$, bottom to top.
  Straight lines have slopes as in the legend. (d) Correlation length vs.\
  time. Straight lines have slopes 1/2 (lower left corner) and 2/3 (upper right corner).
  All units are arbitrary.}
\label{fig:3}
\end{figure}

Given the relevance of fluctuations in these dynamics, we assess
them in Fig.\ \ref{fig:3} (a), where we show the time evolution of the
global interface roughness, $W_{\rm CM}(t)= \langle(P_i(t)-R_{\rm
  CM}(t))^2\rangle^{1/2}$, with $R_{\rm CM}(t)$ being the position of
the CM \cite{note:CM}. Scaling behavior $W(t)
\sim t^{\beta}$ holds, with $\beta \simeq 1/3$, {\em both} in regimes
I and II. As standard for kinetic roughening systems in a circular
geometry, $W(t)$ does not saturate due to the non-interrupted growth
of the system size \cite{pastor:2007}. Moreover, as indicated in
Fig.\ \ref{fig:2} (c), during regime II the quadrupole moment $Q$ decays
as $t^{-2/3}$, which follows if we estimate $Q$ as the ratio of the
radial fluctuations to the average radius, $t^{\beta-1}\approx
t^{-2/3}$. We also measure the local roughness $w(l,t)$, namely, the
interface fluctuations (restricted to windows of size $l$) around a
fitting circular arc, which is drawn with respect to the CM. Data are
shown in Fig.\ \ref{fig:3} (c) as functions of $l$, for several
times. Scaling behavior ensues, $w(l) \sim l^{\alpha}$, provided that,
as in the standard Family-Vicsek (FV) Ansatz \cite{barabasi:book}, the
window size $l$ is smaller than a correlation length $\xi(t)$, which
itself grows as $\xi(t) \sim t^{1/z}$. The FV scaling relation $z =
\alpha/\beta$ holds for exponent values $(\alpha,z)$ which are
$(2/3,2)$ in regime I, and $(1/2,3/2)$ in regime II, see
Fig.\ \ref{fig:3} (d). Indeed, in both cases $\beta=1/3$, as implied by
$W(t)$. Hence, the fractal dimension, $D_F = 2-\alpha$
\cite{barabasi:book}, changes from 4/3 in regime I to 3/2 in regime
II. Overall, the evolution is from kinetic roughening in the SAW class
(regime I), for which $\alpha_{\rm SAW}=2/3$ \cite{des_cloizeaux:book}
and $\beta_{\rm SAW}=1/3$ \cite{stauffer:1987,devillard:1988}, to
asymptotic KPZ scaling in regime II, for which $\alpha_{\rm KPZ}=1/2$
and $\beta_{\rm KPZ}=1/3$ \cite{barabasi:book}. If the noise amplitude
decreases significantly ($A_n\lesssim 0.01$), the roughness
remains constant in regime I, namely the SAW stage disappears,
the only measurable scaling behavior being the KPZ asymptotics in regime II,
as in the experiments for circular geometry \cite{takeuchi_et_al}.
See Fig.\ \ref{fig:3} (b).

\begin{figure}[!t]
\centering
\includegraphics[angle=270,width=0.5\textwidth]{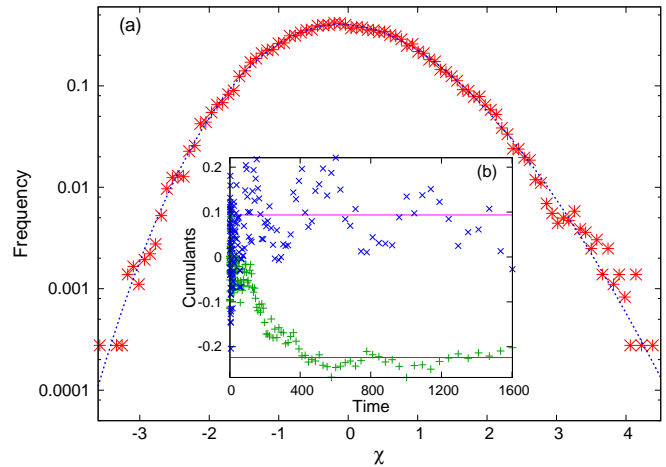}
\caption{(Color online) (a) Histogram of rescaled radial fluctuations $\chi$ at late
  times $t\in[700,1500]$ for the case shown in
  Fig.\ \ref{fig:1}. Simulation data ($*$) and analytic TW-GUE
  distribution (dashed line). (b) Time evolution of skewness
  (lower, $+$) and kurtosis (upper, $\times$) of $R(t)$ for numerical data,
  with the analytic values for the TW-GUE distribution as solid lines.
  Due to a negative $\Gamma$ in Eq.\ \eqref{radius.ev}, the skewness converges
  to {\em minus} the value for the TW-GUE distribution. All units are arbitrary.}
\label{fig:4}
\end{figure}

The progressive dominance of radial fluctuations can be appreciated in
Fig.\ \ref{fig:3} (a), where we plot the time evolution
for the standard deviation of the random variable $R(t)$. Although
this quantity grows fast with $t$, numerically it remains much smaller
than $W(t)$ until onset of regime II, after which both remain
proportional. Actually, we can further explore radial fluctuations in
the asymptotic KPZ regime. Thus, following Pr\"ahofer and Spohn
\cite{praehofer:2000}, we rewrite
\begin{equation}
R(t)\approx \rho_0 + V t + \Gamma t^\beta \chi ,
\label{radius.ev}
\end{equation}
where $\rho_0$, $V$ and $\Gamma$ are constants, $\beta=\beta_{\rm
  KPZ}$, and $\chi$ is a random variable with zero average and unit
variance, whose probability distribution is stationary and corresponds
to the GUE-TW distribution
\cite{praehofer:2000,sasamoto:2010_etal,takeuchi_et_al,yunker:2013,nicoli:2013}. We
have collected the instantaneous radii data for 17 different times in
the range $t \in [700,1500]$, i.e.\ well within regime II, for 2500
noise realizations, in order to check this conjecture. Results are
shown in Fig.\ \ref{fig:4} (a), where we plot the probability distribution
of $\chi$, obtained following the procedure described in
\cite{note:Gamma}, and compare it with the analytical result
\cite{corwin:2012}.
Moreover, we have measured the third and fourth cumulants of this
$\chi$ distribution, i.e.\ the skewness and kurtosis, which are,
respectively, $\langle\chi^3\rangle_c\approx 0.233$ and
$\langle\chi^4\rangle_c\approx 0.0733$. The theoretical values for the
TW-GUE distribution are \cite{praehofer:2000} $0.224$ for the skewness
and $0.093$ for the kurtosis, which are close enough. For comparison
\cite{praehofer:2000}, for the TW-GOE distribution the skewness is
$0.293$ and the kurtosis $0.165$, both being zero for the Gaussian
distribution. Fig.\ \ref{fig:4} (b) shows the time evolution of the
cumulants of $R(t)$ towards the TW-GUE values. We must remark the {\em
  negative} sign that we obtain for parameter $\Gamma$ in
Eq.\ \eqref{radius.ev}, implying a negative skewness for
$R(t)$. Physically, this is due to the fact (data not shown) that, in
regime II, the number of cavities removed per unit length and unit
time by the self-intersection removal condition is smaller than the
number of removed outgrowths.

In summary, while for relatively small noise, perhaps as the experimentally
studied case \cite{takeuchi_et_al}, only KPZ scaling is obtained, for large noise
intensities Eq.\ \eqref{ckpz} predicts a circular
interface to cross over in time between an early-time SAW regime, in which it behaves
as a freely fluctuating ring ``polymer'', and the late-time regime
controlled by KPZ fluctuations in the presence of non-zero average
curvature. For small times, the local driving does not suffice
to counteract fluctuations, so that the average circular shape smears out,
interactions among interface points being controlled by
surface tension (note the dynamic exponent indeed is $z=2$ in
regime I). Since $\xi(t)$ increases while $R(t)$ stays almost constant,
eventually the system becomes fully correlated. From that time on, the increasing length
needs to be accommodated in the finite area enclosed by the
initial radius, and the intersections removal mechanism becomes
relevant, smoothing out the interface. Because of the average (convex) circular
geometry, cavities are removed more frequently than outgrowths, and
the interface starts to grow leading to the expected KPZ regime, with TW-GUE characteristics.

Our results conspicuously connect the celebrated 2D SAW and KPZ universality classes,
both of which underscore the importance of geometrical constraints for scaling behavior,
in and out-of-equilibrium. Crucially, the transition in time between them can only be elucidated through
the existence of overhangs, which eludes other {\em continuum} models
of kinetic roughening. Hence, the large noise regime I of
Eq.\ \eqref{ckpz} might constitute a scaling limit for 2D ring
SAW \cite{cardy_et_al}, while providing an efficient algorithmic
procedure to generate them \cite{janse:2009}. Alternative
connections between the KPZ and SAW classes are available, namely,
between iso-height lines of the 2+1 dimensional (3D) KPZ equation and 2D SAW-related formulations
\cite{saberi_et_al}. In general, the conformation and dynamics of
circular polymers is still a subject of considerable interest
\cite{robertson:2006}. Current experimental capabilities reach even
down to single molecule experiments \cite{witz:2011}, so that one
might speculate on the possibility to observe a dynamical transition
of the type elucidated here in appropriate non-equilibrium, 2D
constrained settings \cite{oh:2012}.

\begin{acknowledgments}
We thank M. Castro, A.\ Celi, M. Nicoli, and T.\ Lagatta for very
useful discussions. This work has been partially supported through
grant 
FIS2012-38866-C05-01 (MINECO, Spain).
\end{acknowledgments}

\begin{acknowledgments}
We thank M. Castro, A.\ Celi, M. Nicoli, and T.\ Lagatta for very
useful discussions. This work has been partially supported through
grant 
FIS2012-38866-C05-01 (MINECO, Spain).
\end{acknowledgments}

\section{Supplemental material}
 
 \subsection{The covariant KPZ equation}

The Kardar-Parisi-Zhang (KPZ) equation describes the dynamics of the
interface between a 2D unstable phase and a stable phase \cite{kardar:1986}.
The interface is represented as the graph of a
single-valued function $h(x,t)$, i.e.: a {\em height} field, in the
small-slopes approximation. Its dynamics follows a non-linear
stochastic partial differential equation (eq. (1) of the main text):
\begin{equation}
\partial_t h = v + \nu \partial^2_x h + {\lambda\over 2} \left( \partial_x h\right)^2 + \sqrt{D} \, \eta
\label{kpz}
\end{equation}
where $v$, $\lambda$, $\nu$ and $D$ are parameters quantifying (average
and excess-) growth, surface tension, and noise intensity, with $\eta$ being a
random field with unit variance, white in space and time. The study of the KPZ equation
has allowed to elucidate the {\em KPZ universality class}, which encompasses many
different models, both discrete and continuous.


The KPZ equation relies on approximations including the assumption of
small slopes and the absence of overhangs. Moreover, by construction it
cannot describe explicitly systems, like circular interfaces, which cannot
be specified by the graph of a single-valued function. There have been attempts
to study an intrinsic-geometry version of the KPZ equation, free from such
assumptions, which have found difficulties due to the strong non-linearity
of the equation; on the other hand, phase-field studies are too expensive
computationally to provide a thorough analysis of the scaling
properties, see a brief overview in \cite{rodriguez-laguna:2011}.

A solution to all these shortcomings was proposed in
\cite{rodriguez-laguna:2011}, in which a {\em covariant form} of the
KPZ equation is put forward (eq. (2) of the main text):
\begin{equation}
\pl_t \mathbf{r} = \( A_0 + A_1 K(\mathbf{r}) + A_{n} \eta(\mathbf{r},t) \) \mathbf{u}_n.
\label{ckpz}
\end{equation}
where $A_0$ and $A_1$ refer to growth speed and surface
tension, while $A_n$ provides the noise
amplitude.
Here, $K(\mathbf{r})$ stands for the local extrinsic curvature and
$\mathbf{u}_n$ is the local normal direction, pointing towards the
non-aggregated phase. The long time behavior of equation (\ref{ckpz}) in band
geometry with periodic boundary conditions was studied in
\cite{rodriguez-laguna:2011}, where ample numerical evidence was given that it
falls into the KPZ universality class, with $\alpha=1/2$ and
$\beta=1/3$.

The main advantage of equation (\ref{ckpz}) is that it can be studied in
circular geometry directly as it stands. Its covariance properties are even
stronger, since any change in the background metric can be
straightforwardly absorbed by the equation.

In order to obtain a well-defined dynamical system for an interface,
equation (\ref{ckpz}) must be complemented with a prescription for the
treatment of intersections. As in \cite{rodriguez-laguna:2011} and
\cite{blinnikov:1996}, whenever a self-intersection appears in the
interface, we have chosen to {\em remove the smaller component},
whether it is a cavity or an outgrowth. Mathematically, intersection removal
is thus the price one needs to pay in order to have a simply connected interface.
Physically, by comparison with more realistic (e.g.\ stochastic moving boundary)
KPZ-related growth models, of which equation (\ref{ckpz}) and similar models
are effective descriptions, its role for the morphological evolution seems to be quite
marginal even for parameter conditions in which voids and bubbles occur,
see e.g.\ \cite{nicoli:2009}. Indeed, with respect to the kinetic roughening behavior,
the main role in the morphological evolution is played by the envelope of the
so-called active growth zone \cite{barabasi:book}, that corresponds in our case to the simply
connected interface that we keep track of after self-intersection removal. Our choice is
akin to the standard procedure by which the full interface dynamics of, e.g.,
discrete growth models that lead to bubbles and overhangs due to bulk vacancies
\cite{krug:1997} is traded for that of such an envelope.

\subsection{Simulation technique}

Most approaches to continuous models of circular growth describe the
interface via a (single-valued) polar function $R(\theta)$, thus
preventing the creation of {\em radial} overhangs. Our approach does not
suffer from this constraint either \cite{rodriguez-laguna:2011}.

Our numerical method of simulation will be the same as in
\cite{rodriguez-laguna:2011}. The interface is represented by a chain of points
(a ``{\em polymer}''), $\{P_i\}_{i=1}^N$, defining a piecewise continuous
curve which always leaves the solid region on its left. The distance
between them is forced to remain in a certain interval $[l_{min},
  l_{max}]$, which is achieved by inserting or removing points
dynamically. All the geometric elements are obtained in a natural
way. Thus, we define the tangent line at $P_i$ as the segment joining
$P_{i-1}$ and $P_{i+1}$, and the normal vector is the unit vector
orthogonal to it pointing outwards. The extrinsic curvature at $P_i$
is defined as the inverse of the radius of the circumference which
passes through $P_{i-1}$, $P_i$ and $P_{i+1}$, which is a discrete
approximation to the osculating circle. This extrinsic curvature is
signed: it is defined to be positive when $P_i$ is at the left side of
the tangent line. In other terms, when the interface is locally
convex.

Numerical integration of the stochastic partial differential equation
was performed with an Euler-Maruyama algorithm. In all cases, the initial
condition will be a circular interface with small radius $R_0$, centered at
the origin $(0,0)$.

Our numerical scheme does not suffer from instabilities from the non-linear
character of the equation. An important aspect of the algorithm is the
detection of self-intersections, for which we have employed the
technique described in \cite{rodriguez-laguna:2011}. Our self-intersection
removal algorithm works in three stages: (1) mark the pairs of
segments which are candidates for intersection by evaluating whether
the cartesian boxes which contain them overlap, (2) for the candidate
pairs, find out whether they do possess an intersection point, (3) if
the intersection point exists, mark it as a new point of the
interface, detect the smaller component (in length) and remove it.

For the results described in the main text, we have run 2500 samples
of equation (\ref{ckpz}), starting with a circle with $R_0=10$.  The
values of the parameters are $A_0=0.01$, $A_1=0.01$ and $A_n=1$, with
$T_{max}=1600$, $\Delta t=0.1$, $l_{min}=0.1$ and $l_{max}=1$. Results
have been checked to ensure the continuum limit has been suitably
approximated. A small number of samples (50) were allowed to continue
up to $T_{max}=16000$.

\subsection{Measurement procedure}

Let us consider the set of points $\{P_i\}_1^N$ that determines the
interface at a given time. There are several possible approaches in
order to measure the roughness of the interface:

\begin{itemize}
\item{} Fit to a circumference {\em with center at the origin}. The
  average radius will be given by $R^2_0\equiv\<|P_i-(0,0)|^2\>$,
  where $\<\cdot\>$ denotes average over all realizations of the
  interface for a given time. The roughness $W$ is given by the
  average deviation from the fitting circumference:
  $W_0\equiv\<(R_i-R_0)^2\>^{1/2}$, i.e.: the error of this fit.

\item{} Fit to a circumference with {\em adjustable center}, which we
  may interpret as the center of mass (CM). The average radius,
  $R_{CM}$ will always be smaller than $R_0$. The roughness is again
  given by $W_{CM}\equiv \<(R_i-R_{CM})^2\>^{1/2}$, and it is smaller
  than in the fixed-center case in all cases.

\item{} The best shape to fit the interface {\em need not} be a
  circle. We may also try a fit to an {\em ellipse}. In this case, the
  output of the fitting procedure is the center of mass, the two
  principal radii, $R_a$ and $R_b$ (the average radius corresponding
  to $R=\sqrt{R_aR_b}$) and the angle which the largest radius makes
  with a fixed direction $\theta$.
\end{itemize}

Since this last fit to ellipses is extremely expensive from a
computational point of view, we have established the following
compromise:

\begin{itemize}
\item{} Average radius from the origin $R_0$ and from the center of mass,
$R_{CM}$.

\item{} Roughness, both from the origin and from the center of mass, $W_0$
and $W_{CM}$.

\item{} Root-mean-square distance of the CM from the origin, $D_{CM}$.

\item{} Anisotropy of the interface: we measure the {\em quadrupole
  moment}:
\begin{equation}
Q^2={\< |x^2-y^2|\> \over \<r^2\>},
\label{quadrupole}
\end{equation}
where $x$, $y$ and $r$ are always relative to the CM. For a
completely isotropic model, like ours, this is a measure of the
eccentricity of the fitting ellipse. Additionally, by defining the
{\em gyration tensor} \cite{family:1985,camacho:1990}
\begin{equation}
S_{ij}=\< x_i x_j\>,
\label{quadrupole.tensor}
\end{equation}
again with $x_i$ and $x_j$ defined with respect to the CM,
another measurement that we have performed is the ratio of its
smallest to largest eigenvalues, $S^2_{G1}, S^2_{G2}$, namely, the
so-called asymmetry parameter \cite{family:1985}
\begin{equation}
\Sigma = \langle S^2_{G1}/S^2_{G2}\rangle .
\label{asymm_param}
\end{equation}

\end{itemize}

All these values are referred to a single snapshot. A second averaging
procedure is required over all the ensemble of realizations of the
solution to the stochastic differential equation (\ref{ckpz}).

A different problem is posed by the {\em morphology curves}, which
show the roughness at each scale, $w(l)$. In \cite{rodriguez-laguna:2011} a
technique was developed to obtain the roughness at a given length, by
using movable windows of width $l$. All points within a window are
fitted to a straight line, and their deviations from that line
constitute the local roughness. In order to transfer this scheme to
the circular geometry setup, we have to take into account that the
natural fit is not to a straight line, but to an arc of
circumference. This problem can be circumvented using the following
``rectification'' procedure:

\begin{itemize}
\item{} For every point $P_i=(x_i,y_i)$, the polar coordinates are found
from the CM:
$P_i=(x_{CM}+R_i\cos(\theta_i),y_{CM}+R_i\sin(\theta_i))$.

\item{} We apply a {\em rectification mapping}: $P_i \to \tilde P_i
=(R_i\theta_i, R_i-R_{CM})$. In other terms: each point is projected
on the fitting circle. The new $x$ coordinate is the arc-length in
this circle. The new $y$ coordinate is its distance to the fitting
circle.
\end{itemize}

Once the new points are found, $\{\tilde P_i\}$, the previous
algorithms apply in a straightforward way.

\subsection{Effective dynamics of the average radius}

We next write down an approximate model for the time evolution of the
average interface radius $R(t)$.  We first integrate Eq.\ \eqref{ckpz}
over the full length of the evolving curve at a fixed time
\begin{equation}
\oint (\partial_t\mathbf{r} \cdot \mathbf{u}_n) \, {\rm d}s = A_0 \, L - A_1 {\cal K} + H ,
\label{int_vn}
\end{equation}
where $L(t)$ is the total curve length, ${\cal K}(t)$ its total
curvature, and $H(t)$ is the integral of the noise contribution. Next
we average \eqref{int_vn} over realizations of the zero-average noise
$\eta$, to obtain
\begin{equation}
2 \pi R \, \frac{{\rm d}R}{{\rm d}t} = 2 \pi A_0 R - 2\pi A_1 ,
\label{mf_vn}
\end{equation}
where we have made additional assumptions:
\begin{itemize}
\item We simplify $\partial_t\mathbf{r} \cdot \mathbf{u}_n = {\rm d}R/{\rm d}t$. Hence, we are neglecting local changes in the normal velocity along the curve.
\item In principle, the factor multiplying $A_0$ in the first term on the right hand side woule be $\langle L \rangle$. We approximate it by the circle length in order to have a differential equation which is closed in $R(t)$.
\end{itemize}
\begin{figure}[t]
\begin{center}
\includegraphics[angle=-90,width=0.5\textwidth,clip=]{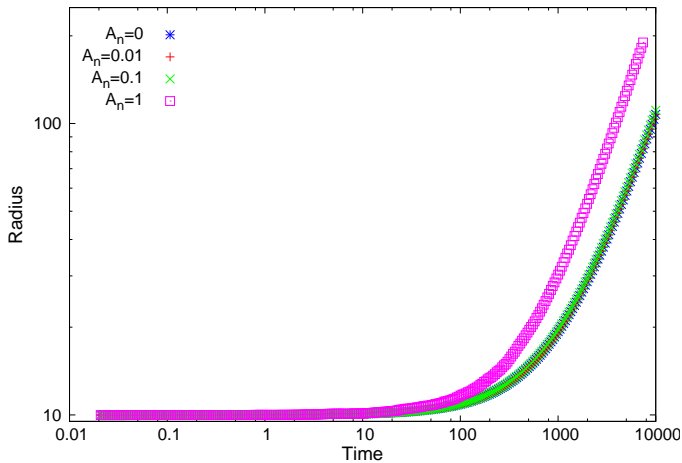}
\end{center}
\caption{Numerical evolution of $R(t)$ according to Eq.\ \eqref{ckpz} for the same parameter choice as discussed in the main text and decreasing noise amplitudes, top to bottom, as in the legend. The uppermost data set is the same as in Fig.\ 2 of the main text (upper panel), which fits Eq.\ \eqref{sol_mf_R} for $\tilde{A}_0 = 0.026$ and $\tilde{A}_1=0.1$.}
\label{fig:1_sm}
\end{figure}
Thus we finally get
\begin{equation}
\frac{{\rm d}R(t)}{{\rm d} t} = \tilde{A}_0 - \frac{\tilde{A}_1}{R(t)} ,
\label{R_mf}
\end{equation}
which is equation (3) of the main text, provided $\tilde{A}_0 = A_0$ and $\tilde{A}_1=A_1$. Note, an equation of the latter form is the simplest one would expect starting from \eqref{ckpz}. Thus, e.g.\ $A_1/R(t)=A_1 K(t)$ for the deterministic case of a circle. Note this surface-tension term contributes to the global interface velocity, in contrast with the case of band geometry. Eq.\ \eqref{R_mf} can be exactly solved as an implicit function for $R(t)$, which reads
\begin{equation}
R(t) - a \ln \left(\frac{R(t)+a}{R_0+a} \right) = R_0 + \tilde{A}_0 \, t ,
\label{sol_mf_R}
\end{equation}
where $R(t=0) = R_0$ and $a=\tilde{A}_1/\tilde{A}_0$. This is the exact solution employed in the fit shown in the main text. As discussed there, the excellent fit suggests equation as an accurate description for $R(t)$, provided one allows for
parameter renormalization whereby $\tilde{A}_0 \neq A_0$ and $\tilde{A}_1\neq A_1$ in general. Actually, such a change
in parameter values is induced by the noise. Thus, in Fig.\ \ref{fig:1_sm} we see that, for sufficiently small but non-zero noise amplitudes $A_n < 1$, the evolution of the radius $R(t)$ cannot be distinguished from the one obtained in the determinstic limit of Eq.\ \eqref{ckpz}.


\begin{references}

\bibitem{witz:2011} G. Witz, K. Rechendorff, J. Adamcik, and
  G. Dietler, Phys. Rev. Lett., {\bf 106}, 248301 (2011).

\bibitem{micheletti:2012} C. Micheletti and E. Orlandini,
  Macromolecules {\bf 45}, 2113 (2012).

\bibitem{luo:2008} K. Luo, T. Ala-Nissila, S.-C. Ying, and
  A. Bhattacharya, Phys. Rev. Lett. {\bf 100}, 058101 (2008).

\bibitem{ercolini:2007} E. Ercolini {\em et al.},
  Phys. Rev. Lett. {\bf 98}, 058102 (2007).

\bibitem{leibler:1987} S. Leibler, R. P. Singh, and M. E. Fisher,
  Phys. Rev. Lett. {\bf 59}, 1989 (1987).

\bibitem{camacho:1990} C. J. Camacho and M. E. Fisher,
  Phys. Rev. Lett. {\bf 65}, 9 (1990).

\bibitem{des_cloizeaux:book} J. des Cloizeaux and G. A. Jannink, {\em
  Polymers in solution, their modelling and structure} (Oxford
  University Press, Oxford, 1990).

\bibitem{cardy_et_al} J. L. Cardy, Adv. Phys. {\bf 318}, (2005);
  M. Bauer and D. Bernard, Phys. Rep. {\bf 432}, 115 (2006);
  I. A. Gruzberg, J. Phys. A: Math. Gen. {\bf 39} 12601 (2006).

\bibitem{takeuchi_et_al} K. A. Takeuchi and M. Sano,
  Phys. Rev. Lett. {\bf 104}, 230601 (2010); K. A. Takeuchi, M. Sano,
  T. Sasamoto, and H. Spohn, Sci. Rep. {\bf 1}, 34 (2011);
  K. A. Takeuchi and M. Sano, J. Stat. Phys. {\bf 147}, 853 (2012); K.A. Takeuchi, 
  Phys. Rev. Lett. {\bf 110}, 210604 (2013).

\bibitem{yunker:2013} P. Yunker {\em et al.}, Phys. Rev. Lett. {\bf
  110}, 035501 (2013); {\em ibid.} {\bf 111}, 209602 (2013).

\bibitem{nicoli:2013} M. Nicoli, R. Cuerno, and M. Castro, Phys. Rev. Lett. {\bf 111}, 209601 (2013).

\bibitem{saito:2012} Y. Saito, M. Dufay, and O. Pierre-Louis,
  Phys. Rev. Lett. {\bf 108}, 245504 (2012).

\bibitem{praehofer:2000} M. Pr\"ahofer and H. Spohn,
  Phys. Rev. Lett. {\bf 84}, 4882 (2000); Physica A {\bf 279}, 342
  (2000).

\bibitem{barabasi:book} A.-L. Barab\'asi and H. E. Stanley, {\em
  Fractal Concepts in Surface Growth} (Cambridge University Press,
  Cambridge, UK, 1995).

\bibitem{krug:1997} J. Krug, Adv. Phys. {\bf 46}, 139 (1997).

\bibitem{kardar:1986} M. Kardar, G. Parisi, and Y.-C. Zhang,
  Phys. Rev. Lett. {\bf 56}, 889 (1986).

\bibitem{corwin:2012} I. Corwin, Random Matrices: Theor. Appl. {\bf
  1}, 1130001 (2012).

\bibitem{calabrese} T. Gueudre,  P. Le Doussal, A. Rosso, A. Henry, and P. Calabrese, 
Phys. Rev. E {\bf 86}, 041151 (2012).


\bibitem{sasamoto:2010_etal} T. Sasamoto and H. Spohn,
  Phys. Rev. Lett. {\bf 104}, 230602 (2010); G. Amir, I. Corwin, and
  J. Quastel, Commun. Pure Appl. Math. {\bf 64}, 466 (2011).

\bibitem{kriecherbauer:2010} T. Kriecherbauer and J. Krug, J. Phys. A:
  Math. Theor. {\bf 43}, 403001 (2010).


\bibitem{alves:2011_etal} S. G. Alves, T. J. Oliveira, and
  S. C. Ferreira, EPL {\bf 96}, 48003 (2011); T. J. Oliveira,
  S. C. Ferreira, and S. G. Alves, Phys. Rev. E {\bf 85}, 010601(R)
  (2012).

\bibitem{rodriguez-laguna:2011} J. Rodr\'{\i}guez-Laguna,
  S. N. Santalla, and R. Cuerno, J. Stat. Mech.: Theor. Exp. (2011)
  P05032.

\bibitem{el_alaoui_faris:09} M. D. El Alaoui Faris {\em et al.} Phys. Rev. Lett.
{\bf 102}, 038102 (2009).

\bibitem{blinnikov:1996} S. Iv. Blinnikov and P. V. Sasorov,
  Phys. Rev. E {\bf 53}, 4827 (1996).

\bibitem{sm} See supplemental material for further details on
Eqs.\ (6), (7), and the simulation and measurement procedures
that we have employed.

\bibitem{family:1985} F. Family, T. Vicsek, and P. Meakin,
  Phys. Rev. Lett. {\bf 55}, 641 (1985).


\bibitem{bishop:1986} M. Bishop and C. J. Saltiel, J. Chem. Phys. {\bf
  85}, 6728 (1986).

\bibitem{alim:2007} K. Alim and E. Frey, Phys. Rev. Lett. {\bf 99},
  198102 (2007).


\bibitem{note:CM} We have verified that ---in contrast with cases like
  the Eden model, see S. C. Ferreira Jr. and S. G. Alves,
  J. Stat. Mech. P11007 (2006)--- measurements of the roughness with
  respect to a fixed point instead of the CM do not change results on
  critical exponent values.

\bibitem{pastor:2007} J. M. Pastor and J. Galeano, Central
  Eur. J. Phys. {\bf 5}, 539 (2007).

\bibitem{stauffer:1987} D. Stauffer and N. Jan, Can. J. Phys. {\bf 66}, 187 (1988).

\bibitem{devillard:1988} P. Devillard, Physica A {\bf 153}, 189 (1988).

\bibitem{note:Gamma} To obtain the histogram for $\chi$, we follow a
  procedure similar to Ref.\ \cite{takeuchi_et_al}. Namely, the raw
  $(t,R)$ pairs are fit to a linear function of time, providing
  $\rho_0$ and $V$. The squared residuals are fit to a power law of
  time, yielding $|\Gamma|$ and $\beta\approx 0.3353$. From here we
  estimate the values $\chi=(R-\rho_0-Vt)/(\Gamma t^\beta)$. Finally,
  $\Gamma$ has the same sign as the skewness of $R(t)$.

\bibitem{janse:2009} E. J. Janse van Rensburg, J. Phys. A: Math. Theor. {\bf 42}, 323001 (2009).

\bibitem{saberi_et_al} A. A. Saberi and S. Rouhani, Phys. Rev. E {\bf
  79}, 036102 (2009).

\bibitem{robertson:2006} R. M. Robertson, S. Laib, and D. E. Smith,
  Proc. Natl. Acad. Sci. USA {\bf 103}, 7310 (2006).

\bibitem{oh:2012} Y. Oh, H. W. Cho, J. Kim, C. H. Park, and B. J. Sung,
Bull. Korean Chem. Soc. {\bf 33}, 975 (2012).

\bibitem{nicoli:2009} M. Nicoli, M. Castro, and R. Cuerno, J. Stat. Mech.: Theor. Exp. (2009) P02036.

\end{references}
\end{document}